\documentclass[usenatbib,usegraphicx,twocolumn]{mn2e}
\def\pasj{PASJ}
\def\mnras{MNRAS}
\def\apj{Ap.J}
\def\C{{\rm S}}
\def\N{{\rm N}}

\def\u{{\bf U}}

\def\HI{{\rm HI}}
\def\H{{\rm H}}

\def\A10{A_{10}}

\def\rn{r_{\nu}}
\def\rnp{r_{\nu}^{'}}

\def\k{{\bf k}}
\def\kp{k_\parallel}

\def\m{{\bf m}}

\def\k{{\bf k}}

\def\u{{\bf U}}

\def\kop{{k_{1 \parallel}}}
\def\ktp{{k_{2 \parallel}}}

\def\HI{{\rm HI}}
\def\H{{\rm H}}

\usepackage{epsfig}
\usepackage{graphics}
\begin{document}
  \title[Probing the bispectrum at high redshifts.] 
	{Probing the bispectrum  at high redshifts using 21 cm
	  HI  observations }
	\author[S. S. Ali, S. Bharadwaj and S. K. Pandey ]{SK. Saiyad
  Ali$^{1}$\thanks{Email:saiyad@cts.iitkgp.ernet.in}, Somnath
	  Bharadwaj$^{1}$\thanks{Email:somnath@cts.iitkgp.ernet.in}
	and Sanjay K. Pandey$^{2}$\thanks{Email:spandey@iucaa.ernet.in}
\\ $^{1}$ Department of Physics and Meteorology \&
	Centre for Theoretical Studies , IIT Kharagpur,  Pin: 721 302 , 
	India. 
\\$^2$ Deptt. of Mathematics ,L.B.S.College, Gonda 271001 , India \& Visiting Associate, Inter University Centre for Astronomy and Astrophysics,\\ Pune , India.}
\maketitle
\begin{abstract} 
Observations of  fluctuations in the redshifted  21
cm radiation from neutral hydrogen (HI) are  perceived to be an  
important future probe of the universe at high redshifts. Under the
assumption that at redshifts $z \le 6$ (Post-Reionization Era)  the HI
traces the underlying dark matter with a possible bias, we investigate
the possibility of using observations of redshifted 21 cm radiation to
detect  the bispectrum arising from non-linear gravitational
clustering and from non-linear bias.  We find  that the expected signal
is $\sim 0.1 {\rm mJy}$ at $325 \, {\rm MHz}$ $(z=3.4)$ for the small
baselines at the GMRT, the strength being a few times larger at higher
frequencies   ($610 \, {\rm MHz},  z=1.3$). Further, the magnitude of
the signal from the bispectrum is predicted to be comparable to that
from the power spectrum, allowing a detection of both in roughly the
same integration time.  The HI signal is
found to be uncorrelated beyond frequency separations of $\sim 1.3
{\rm MHz}$ whereas the continuum sources of continuum are expected to
be correlated  across much larger frequencies.  
 This signature can in   principle be used to distinguish the HI
 signal from the contamination. We also consider the possibility of
 using observations of the bispectrum to determine  the linear and
 quadratic bias parameters of the HI at  high redshifts, this having
 possible implications for theories of galaxy formation. 
\end{abstract}


\section{Introduction}
Quantifying non-Gaussian features in the large-scale matter
distribution is an important issue in cosmology. It is widely accepted
that the large-scale structures originated from  initially small
density  fluctuations which were a Gaussian random field.
Non-Gaussian features arise as the fluctuations  grow through the
process 
of gravitational instability (eg. \citealt{peebles1}), and 
quantifying these at various stages of the growth is expected to
yield a significant amount of information. The three-point correlation
function and its   
Fourier transform, the bispectrum, are the lowest order statistics
sensitive to non-Gaussian features.   

The determination of the three-point correlation function was
pioneered by \citet{peebgroth} using the galaxies in the Lick and
Zwicky angular catalogues. Later studies of the three point
correlation function using galaxy  surveys include 
\citet{peebles2}, \citet{bean}, \citet{efst}, \citet{hale},  
 \citet{Jing1}, \citet{gaz94}, \citet{frmn94}, \citet{Jing2}. The
 recently completed Two -degree Field Galaxies Redshift Survey (2dfGRS) and
the currently   ongoing  SDSS have been used to measure the galaxy
three-point correlation function at an unprecedented level of accuracy
(2dFGRS, \citealt{Jing3}; \citealt{kayo}; \citealt{gaz05}; SDSS )
allowing various 
issues like the luminosity and colour dependence to be
addressed. There have recently been a few measurements of the
bispectrum for the IRAS galaxies (\citealt{feldman};  \citealt{sco4})
and for the 2dFGRS \citep{verde2002} which have (as discussed later)
been used to estimate the galaxy bias.

All the measurements of the three-point correlation function or the
 bispectrum mentioned here are restricted to low redshifts
 $(z<1)$. This limitation arises because they  are based on galaxy
 surveys which do not extend to very high redshifts. For example  the
 2dFGRS which is the largest completed redshift survey extends to a
 maximum   redshift of around $z=0.3$. Further, a galaxy or quasar
 survey extending out to large redshifts ($z>1$) would select
 only the  brightest objects which would be sparse in number. These
 objects would be  distributed  in the densest regions which are known
 to have a highly biased distribution and their distribution  
would not be representative of the  underlying matter distribution. 
In this  paper we discuss how observations of the redshifted 
 $21$~cm  emission from neutral hydrogen (HI)  can be used to measure
 the bi-spectrum at high   redshifts ({\it ie.} $z \sim 1$ and
 higher).  

The HI density in the redshift range $1.7 - 5.5$ is known from
observations of Ly-$\alpha$ absorption lines along lines of sight to
quasars (\citealt{lombardi1}; \citealt{lombardi2}; \citealt{peroux}; 
\citealt{prochaska}). These observations indicate that $\Omega_g(z)$, 
the comoving density of neutral gas expressed as a fraction of the
current critical density, has a value $\sim 10^{-3}$. Recent 
observations \citep{prochaska} show that $\Omega_g(z)$ drops by about
$25 \%$  at $z =2.2- 2.5$. Current observations of  $\Omega_g(z)$ are 
consistent with no evolution at higher redshifts ($ 3 \le z \le 5.5$).
The bulk ($ \sim 70 \%$) of the HI resides in high column-density
regions $(\ge 2. \times 10^{20}\, \rm cm^{-2})$  which are possibly the
progenitors of present day galaxies.  The redshifted 21 cm ($1420\, {\rm
  MHz}$) emission  from this HI is present  as a background radiation
in radio observations at all frequencies below $1420\, {\rm MHz}$.  The
fluctuations in this background radiation arise from  fluctuations in
the HI density and from peculiar velocities at the redshift where the
radiation originated (\citealt{BNS1}, hereafter BNS). The possibility of
detecting this  holds the potential of being an important 
tool to probe the large-scale structures in the redshift range $z = 1 
- 6$. This is particularly significant in the context of the GMRT\citep{swarup}
 which operates in several bands in the frequency range of
interest.  
 It is also  interesting to  note that the fluctuations in the
 background HI 
radiation is expected to exceed the fluctuations in the CMBR by a
factor of $ 10 -100$  over the frequency range of interest (BNS).  

Complex visibilities at different baselines and frequencies are the
primary quantity  measured in radio interferometric
observations. Correlations between pairs of visibilities directly
probe the three dimensional power spectrum of HI fluctuation at the
redshift where the radiation originated.   This has
been studied  in a series of papers (\citealt{BS1}; \citealt{BP3},
\citealt{BS4})  which develop the relation between visibility
correlations 
and the  HI power spectrum, and estimate the expected signal in the
redshift range $z <6$  both analytically and using N-body simulations.     
This formalism has later been extended  (\citealt{BA5})  to quantify
the expected HI visibility correlation  signal over a large redshift
range extending from the dark ages $(z \sim 100) $ to the present
epoch.  The possibility of using visibility correlations to probe the 3D
HI power spectrum has also been considered by \citet{morl} while
\citet{zald} propose the use of the angular power spectrum, both 
in the context of detecting  the Epoch of Reionization (EOR) signal.  

\citet{BP5} ( hereafter BP05) have, in a recent paper, investigated
the possibility of using correlations between visibilities observed at
three different baselines and frequencies to measure the bispectrum of
HI fluctuations at the epoch of reionization. The HI distribution
during the epoch of reionization will  be  determined more by the
size, distribution and topology of the ionized 
regions and it is  anticipated  that this will not trace the
underlying dark matter distribution.  It is thus expected that the
bispectrum of the HI during reionization will tell us more about the
ionization process and less about the non-Gaussianity arising from the
non-linear growth of density perturbation. 
In this paper we investigate how three visibility correlations can be
used to study the HI bispectrum at redshifts $z < 6$, where it is
reasonable to assumed that the  large scales  HI  distribution  traces
the dark matter with a possible bias.  

We next briefly discuss what we can hope to learn from  observations
 of non-Gaussianities in the  HI.  In perturbation theory the lowest
 order  at which there is  a non-zero  three point correlation function  
is the second order (\citealt{fry84}; \citealt{bharadwaj1}). These
predictions are expected to be valid on large  scales which are 
weakly  non-linear, and it is in principle possible to compare these
with observations and test the
 currently accepted scenario for
structure formation. The bispectrum has been perceived to be  a more  
effective  statistics for quantifying non-Gaussianity in the weakly
non-linear regime and its theory has been developed in
\citet{fry94}, \citet{hivon}, \citet{matar}, \citet{verde98},
\citet{sco1},  \citet{sco2} and \citet{sco4}. These investigations
show that it possible to use observations of the bispectrum (or
equivalently the three point correlation function) to
determine the linear and quadratic bias parameters. This has been
actually carried out using various galaxy surveys like the APM
 galaxies  and the IRAS galaxies (\citealt{feldman}; 
\citealt{sco4}).  In a recent analysis \citet{verde2002} have analysed
the bispectrum of the 2dFGRS to conclude $b_1=1.04 \pm 0.11$ for the
linear bias parameter and $b_2=-0.054 \pm 0.08$ for the quadratic bias
parameter, indicating that the 2dFGRS galaxies are an unbiased tracer
of the underlying dark matter. Further, the linear bias parameter was
used in combination with the  redshift distortion 
parameter $\beta \approx \Omega_{m0}^{0.6}/b_1$ measured  from the same
survey   \citep{peacock}
to determine the density of the universe $\Omega_{m0}=0.27
\pm 0.06$. A new analysis of the three point correlation function of
 the 2dFGRS galaxies \citep{gaz05} confirms that the large-scale
 structures formed through  gravitational instability starting from Gaussian
 initial conditions. The study concludes  that the galaxies do not   
trace the underlying dark-matter distribution with $b_1=0.93^{0+.10}_{-0.08}$  
 and $c_2=b_2/B_1=-0.34^{+0.11}_{-0.08}$. Observations of the HI
 bispectrum would allow us to carry out such studies at redshifts
 $z>1$, providing further tests of the gravitational instability
 picture. Further, it would be possible to determine the  bias of the
 HI  at  high redshifts,  this being a potential probe of  galaxy
 formation.    

The cosmological evolution of the neutral hydrogen
 during the Epoch of Reionization $(z=6 - 20)$  and the
 pre-Reionization  era $(z > 20)$, and the expected redshifted 21 cm
 signal are  topics of very intense research (eg.  \citealt{scott};
 \citealt{mmr97};   \citealt{gnedin}; \citealt{shav99};   \citealt{tozz};
  \citealt{iliev};  \citealt{furlanetto}; 
 \citealt{miralda}; \citealt{zald}; \citealt{lz};
 \citealt{bharad5}; \citealt{zar}; \citealt{BA5}; \citealt{BP5}; \citealt{barkana1};
 \citealt{barkana2};  \citealt{barkana3}; \citealt{bow}). We note that most of these papers have no direct bearing on the current work which is  
 restricted to the Post-Reionization Era $(z < 6)$  where it is
 reasonable to assume that the HI traces the dark matter, and the
 process of  gravitational instability has progressed
 sufficiently to produce measurable  departures from the Gaussian
 initial conditions. \citet{subr}, \citet{kum} and \citet{bagla1} have
 considered  the possibility of detecting HI at $z<6$ focusing on the
 possibility of detecting individual features 
 corresponding to big HI clumps. \citet{bagla2} have used N-body simulations
 to estimate the two-point statistics  of the low $z$  HI signal. 

We next present the structure  of the paper.  Section 2 reviews the
formulae needed to calculate the visibility correlation signal
expected from HI at high $z$. In Section 3. we present the results and
discuss their implications.

Finally we note that, unless stated otherwise, we use the values 
$(\Omega_{m_0}, \Omega_{\lambda_0}, \Omega_b {h^2}, h$) = ($0.3, 0.7,
0.02, 0.7)$ for  the cosmological parameters. Further, we use
$\Omega_g=1 \times 10^{-3}$ as a fiducial value and present results
only for this throughout the paper.

\section{Calculating the HI visibility correlation signal}
The quantity  measured in radio-interferometric  observations is the
complex   visibility  $V(\u,\nu)$ which records only the angular
fluctuations  of the specific intensity $I_{\nu}$ on the sky. 
This is  measured for every pair of antennas in a radio-interferometric
array.   For any pair of antennas at a separation ${\bf d}$, we refer
to the two dimensional vector   $\u={\bf  d}/\lambda$ as a baseline. 
In this paper  it has been assumed that all the antennas in the array
are 
coplanar,  and that the  antennas all point vertically
up wards in the direction ${\bf m}$  with ${\bf m \cdot \u}=0$. 
Further, the individual antennas are assumed to have a  Gaussian
beam pattern $A(\theta) =e^{-\theta^2/\theta_0^2}$  with 
$\theta_0 \ll 1$  (in radians) {\it i.e.} the beam width $\theta_0 $ 
of  the antennas is small, and the part of the sky which contributes
to the signal can be  well approximated by a plane.

Fluctuations in the high redshift HI density  contributes to the
visibilities measured at all frequencies below $1420 \, {\rm MHz}$.  
Following BP05 we consider the contribution from the HI signal to
the correlations  
\begin{equation}
\C_2(\u_1,\u_2,\Delta \nu)=\langle V(\u_1,\nu + \Delta \nu) V(\u_2,\nu)
\rangle 
\end{equation}
and 
\begin{eqnarray}
\C_3(\u_1,\u_2,\u_3,\Delta \nu_1,\Delta \nu_2)&=&\langle V(\u_1,\nu +
\Delta \nu_1) V(\u_2,\nonumber \\ &&\nu  + \Delta \nu_2)   V(\u_3,\nu)
\rangle 
\end{eqnarray}
expected between the visibilities at different baselines and
frequencies.  Here the two visibility correlation function
$\C_2(\u_1,\u_2,\Delta \nu)$ denotes the correlation expected between
the visibilities measured
at two different baselines $\u_1$ and $\u_2$  one at the  
frequency $\nu$ and the other at $\nu + \Delta \nu$.  It should be
noted that although we have shown $\C_2$ as an explicit functions of  the 
frequency differences $\Delta \nu$ only, it  also depends  on  the central
value $\nu$  which is not shown as an explicit argument.  The value
of $\nu$ will be clear from the context of the discussion.  
 Further,  throughout our analysis we  assume  that all
 frequency differences are much   smaller than the central frequency
 {\it ie.} $\Delta \nu/\nu \ll 1$.   The definition of the three 
visibility correlation $\C_3$ closely follows that of $\C_2$.

It has been shown that the two-visibility correlation probes the power
 spectrum of HI fluctuations (\citealt{BS1}, \citealt{BA5}) and the
 three visibility correlation the bispectrum  (\citealt{BP5}). 
Considering   $\C_2(\u_1,\u_2,\Delta \nu)$ first, we  note that it has
 a significant value only when $\mid \u_2  +  \u_1 \mid  < (\pi
 \theta_0)^{-1}$.  
Therefore we restrict the analysis to $\u_1=-\u_2 =\u$  which we
 denote as $\C_2(\u,\Delta \nu)$. This is related to the  redshift
 space  HI power  spectrum $P_{\rm HI}(\k)$ as  
\begin{eqnarray}
\C_2(U, \Delta \nu) = 
\frac{\bar{I}^2_{\nu} \theta_0^2}{2 \rn^2} 
 \int_0^{\infty}
d \kp \, P_{\HI}(\k)  
\cos(\kp \rnp \Delta \nu) \,.
\label{eq:a9}
\end{eqnarray}
where $\rn$ is the comoving distance to the HI whose $1420 {\rm \,  MHz}$
radiation is observed at  a frequency $\nu$,  $\rnp=d \rn/d\nu$,  
$\k=k_{\parallel} {\bf m} +  (2 \pi/\rn) \u$ and 
$\bar{I}_{\nu}=2.5 \times 10^2 \, \frac{\rm Jy}{\rm sr}  \,
  \left(\frac{\Omega_b   h^2}{0.02}\right)  \left(\frac{0.7}{h}
  \right) \frac{H_0}{H(z)}$. A point to note is that $P_{\rm HI}(\k)$
  includes the effects of  redshift space distortions, and is
  therefore anisotropic along the  the  line of sight  $\m$. 
 We further  note that $\C_2$  is real and  isotropic {\it ie} it
 does not depend on the direction of the baseline $\u$ whereby  we can 
 write $\C_2(U,\Delta \nu)$. 

Considering the three visibility correlation
$\C_3(\u_1,\u_2,\u_3,\Delta \nu_1,\Delta \nu_2)$ next, this has a
significant value only if  $\mid    \u_1+\u_2+\u_3 
 \mid \le  (\pi \theta_0)^{-1}$. We therefore restrict our analysis
 to only those combinations of baselines for which
$\u_1+\u_2+\u_3=0$. Further $\C_3$ is real and it depends only on the
 shape and size of the triangle which is completely specified by the
 magnitude of the three baselines $(U_1,U_2,U_3)$. We have 

\begin{eqnarray}
  \C_3(U_1,U_2,U_3,\Delta \nu_1,\Delta \nu_2)
 & = &\frac{\bar{I}^3_{\nu} \, \theta_0^2}{12 \, \pi \, \rn^4} 
  \int d \kop  d \ktp 
  \cos[(\kop \,\Delta  \nu_1 \nonumber \\& +& \ktp  
 \Delta \nu_2) \rnp] \, 
  \times  B_{\HI}(\k_1,\k_2,\k_3) \label{eq:s2}
\end{eqnarray}

where $\k_1=\kop \m + (2 \pi/\rn) \u_1$, $\k_2=\ktp \m + (2 \pi/\rn)
\u_2$,  $\k_3=-(\kop + \ktp) \m + (2 \pi/\rn) \u_3$ and $B_{\rm
  HI}(\k_1,\k_2,\k_3)$  is the redshift space HI bispectrum.  

It is next necessary to specify the form of the redshift space HI
power spectrum $P_{\HI}$ and bispectrum $B_{\HI}$. Fluctuations in the
redshifted HI radiation arise from  fluctuations in the HI density
and from HI peculiar velocities, and it is necessary to include both
these effects in $P_{\HI}$ and $B_{\HI}$. 
We assume that the HI traces the dark matter with
a possible bias, and retain terms upto the quadratic order in the
relation between the fluctuations in the HI density $\delta_{\rm HI}$
and the fluctuations in the dark matter density $\delta$ 
\begin{equation}
\delta_{\rm HI}=b_1\delta + \frac{b_2}{2} \delta^2 
\end{equation}
where $b_1$ and $b_2$ are the linear and quadratic bias parameters
 respectively. A non-zero $b_2$ would indicate nonlinear biasing of HI
 distribution with respect to underlying mass distribution. It is also assumed that the peculiar velocities are
 determined by the dark matter which dominates the dynamics. 

The HI power spectrum $P_{\rm HI}(\k)$ is related to the real
 space dark matter power spectrum $P(k)$ as
\begin{equation}
P_{\rm HI}(\k)=b_1^2 \bar{x}_{\HI}^2 (1 + \beta \mu^2)^2 P(k) \, 
\label{eq:s3}
\end{equation}
$\mu$ being  the cosine of the  angle between $\k$ and the line of
sight $\m$, and $\beta = f(\Omega_m)/b_1 \approx \Omega_m^{0.6}/b_1$
is the linear distortion parameter. 
The term $(1 + \beta \mu^2)^2$ in equation
(\ref{eq:s3}) incorporates the effect of the peculiar velocities, and
$\bar{x}_{\HI}$ is the mean hydrogen neutral fraction.
Converting $\Omega_{gas}$ to the mean neutral fraction
$\bar{x}_{\HI}=\bar{\rho}_{\HI}/\bar{\rho}{\H}=\Omega_{gas}/\Omega_b$
gives us $ \bar{x}_{\HI}=50 \Omega_{gas}  h^2 (0.02/\Omega_b h^2)$ or
$\bar{x}_{\HI}=2.45 \times 10^{-2}$. 
  It should be noted that $\mid \delta \mid \ll 1$ and terms higher than
  $\delta^2$ have been dropped  in  the power spectrum.

For the HI bispectrum we assume 
\begin{equation}
  B_{\HI}(\k_1,\k_2,\k_3\,,z)=\bar{x}^3_{\HI}\, B_s(\k_1,\k_2,\k_3\,,z)
  \label{eq:d2}
\end{equation}
where $B_s$ is the redshift space bispectrum calculated by 
\citet{verde98}. We have used the form of $B_s$ given in equation (11)
of \citep{verde98}, the expression being quite lengthy we do not
reproduce it here. We note that $B_s(\k_1,\k_2,\k_3)$ has a quadratic
dependence on the dark matter power spectrum $P(k)$, it depends
on  the bias parameters $b_1$ and $b_2$, and on  the parameters of the 
background cosmological model. The value of  $B_s$
also depends on the shape of the triangle formed by $(\k_1,\k_2,\k_3)$.

The fact that the neutral hydrogen is in discrete clouds  makes a
contribution which we do not include here.  This effect originates
from the fact that the HI emission line from individual clouds  has a
finite width, and the visibility correlation is enhanced when $\Delta
\nu$ is smaller than the line-width of the emission from the individual
clouds. Another important effect not included here is that the
fluctuations become significantly non-linear at low $z$. Both these
effects have been studied for the power spectrum using simulations
\citep{BS4}.

 The simple analytic
treatment adopted here suffices for the purposes of this paper where
the main focus is to estimate the magnitude and the nature of the HI
signal, and to investigate the feasibility of using such observations
to  probe structure formation at high redshifts.

\section{Results and Discussion}
We present results for the two  and three visibility
correlation signal expected  from HI at high $z$. In calculating the
expected signal we have used the GMRT parameters,  but it is
 straight forward to scale the results presented here to make the
 visibility correlation predictions for other radio telescopes.  
The only telescope parameter which enters equations (\ref{eq:a9}) and
(\ref{eq:s2}) is  $\ \theta_0 \ $, which is the beam size of the
individual antenna in the array. Further, it should be noted that $ \
\theta_0  \approx 0.6 \times \theta_{\rm   FWHM} \ $. The value of
$\theta_0$ will depend on the physical dimensions of the antennas and
the wavelength of observation.  For the GMRT $ \ \theta_0=1^{\circ} \
$ at $ \ 325 \, {\rm   MHz}$. We  scale this  using $\theta_0 \propto
\lambda$ to  obtain  $ \ \theta_0 \ $ corresponding to different
observation frequencies. We note that the smallest  baseline at
the GMRT is a little smaller than $U=100$ while the largest is around
$U \sim 10,000$, and we present results for $U$ in the range
$10-10,000$. It may be noted that unless mentioned otherwise ,we use
the values ($b_1$ , $b_2$ )$=$ (1.0 , 0.5) for the bias parameters.

\begin{figure}
\includegraphics[width=105mm]{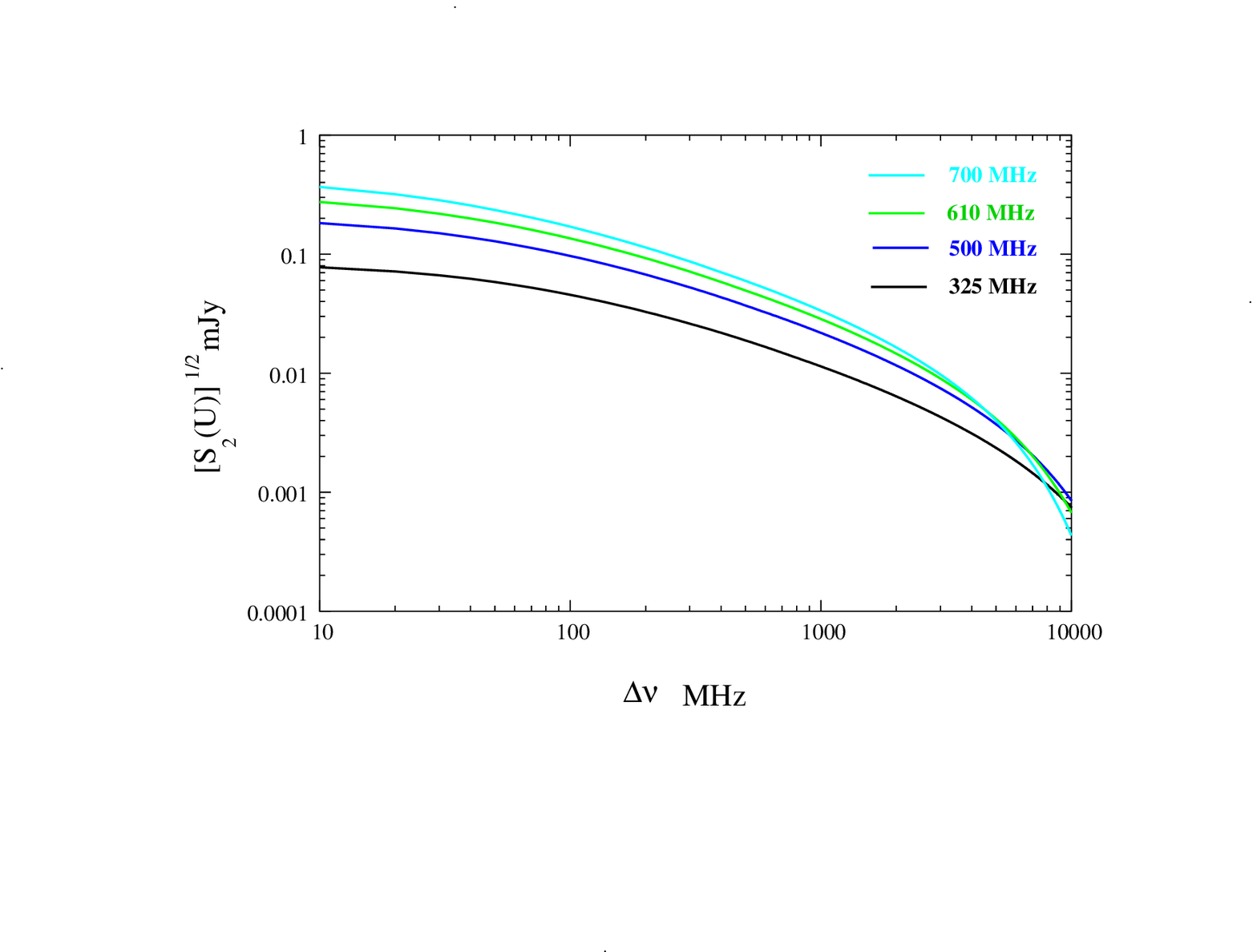}
\caption{This shows the expected  two visibility correlation signal 
  $[\C_2(U,0)]^{1/2}$ as   a function of $ U$  for the different
  values of the central frequency   shown in the figure.}   
\label{fig:1}
\end{figure}

Figure \ref{fig:1} shows the $ [\C_2(U,0)]^{1/2}$, 
the two visibility correlation signal expected when the two
visibilities are at the same baseline and frequency. The signal is
strongest at small baselines, and it is around $0.1 \, {\rm mJy}$ at
$325 \, {\rm MHz}$ {\it ie.} $(z=3.4)$. The signal falls with
increasing $U$, and the $U$ dependence  reflects the
shape of the power spectrum. Further, the signal increases with
increasing central frequency or decreasing redshift,. This is a
consequence  of the fact that the matter power spectrum  grows with
time.

\begin{figure}
  \includegraphics[width=72mm]{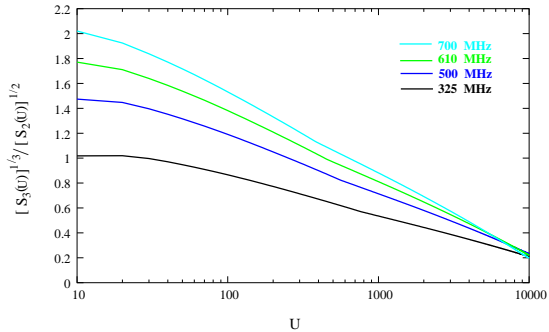}
  \caption{This shows the ratio $\frac {[\C_3(U,0)]^{1/3}}
    {[\C_2(U,0)]^{1/2}}$   as a function of $U$ for 
   the different values of the central frequency shown in
    the figure. }
  \label{fig:2}
\end{figure}

To estimate the magnitude of the expected three visibility correlation
signal $\C_3(U_1,U_2,U_3,\Delta \nu_1,\Delta \nu_2)$, we first
restrict our analysis to  equilateral triangles for which the size of
the baseline $U$  completely specifies the triangle, We further
restrict our analysis to $\Delta \nu_1=\Delta \nu_2=\Delta \nu$  in
which case we can use the notation  $\C_3(U,\Delta\nu)$. 
The ratio  $ \ \frac {[\C_3(U,0)]^{1/3}} {[\C_2(U,0)]^{1/2}}\ $ (Figure \ref{fig:2}) quantifies the relative strength of $\C_3$ as a fraction of $\C_2$. The first point to note is that the three
visibility correlation signal is comparable in magnitude to the two
visibility correlation signal. The ratio of the signal strengths is of
order unity for the small baselines ($U \le 1000$) at $325 \, {\rm MHz}$ and
it is somewhat higher at higher frequencies. Further, $\C_3^{1/3}$
falls faster than $\C_2^{1/2}$  with increasing $U$. It is possible to
understand both these features of $\C_3$ in terms of the fact that the
bispectrum is quadratic in the power spectrum whereby it 
grows faster with time and it has a steeper $k$ dependence at large
$k$.

\begin{figure}
  \includegraphics[width=74mm]{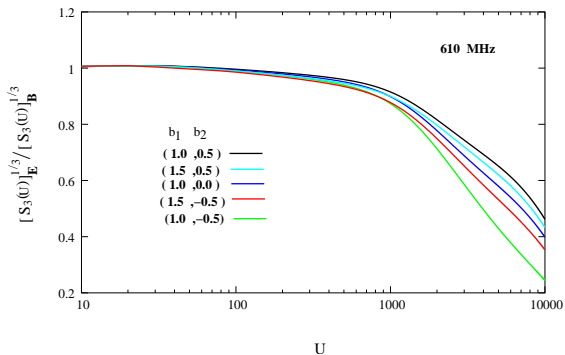}
  \caption{This shows the  ratio  $[\C_3(U,\Delta \nu)]^{1/3}_E/
[\C_3(U,\Delta \nu)]^{1/3}_B$  for $\Delta \nu=0$, where $E$ refers to
equilateral triangles and $B$ to bilateral triangles. Two sides of the
bilateral triangle are $U$ (at $120^{\circ}$) and the third is
$\sqrt{3} U$. The central frequency is $610 {\rm  MHz}$. The results
are for the different combinations  of the bias parameters $(b_1,b_2)$
shown in the figure.}
\label{fig:3}
\end{figure}

The bispectrum $B_s(\k_1,\k_2,\k_3)$ depends on the shape and size of
 the triangle formed by $\k_2,\k_2$ and $\k_3$. This dependence varies
 with the bias parameters $(B_1,b_2)$, and this can be  used to
 observationally determine the values of the  bias parameters.  
To study how the shape and bias dependence  manifests itself in
 $\C_3$, we  have considered  bilateral triangles for which two sides
 are of 
 length $U$ with opening  angle $120 ^{\circ}$ and the third side of
 length $\sqrt{3}U$. We  study the ratio  $[\C_3(U,\Delta
 \nu)]^{1/3}_{\rm E}/[\C_3(U,\Delta \nu)]^{1/3}_{\rm B}$  for $\Delta
 \nu=0$ where the 
 subscripts ${\rm E}$ and ${\rm B}$ refer to equilateral and bilateral
 triangles respectively.   Figure \ref{fig:3} shows the results at 
$610 {\rm MHz}$ $(z=1.3)$, the behaviour at other
 frequencies is similar.  We find that $\C_3$ does not show any
 significant dependence on the triangle shape 
 at  baselines  smaller than $U \sim1000$. There is a shape dependence
 at larger baselines where $\C_3$ decreases faster for  the
 equilateral triangles than for the bilateral triangles. Further,
the ratio has a  distinct dependence on the values of the bias
 parameters. The first point to note is that the ratio is independent
 of $b_1$ for  $b_2=0$. For other values of $b_2$ we find that the
 $U$ dependence  of the ratio  $[\C_3(U,0)]^{1/3}_{\rm
 E}/[\C_3(U,0)]^{1/3}_{\rm B}$   at the large baselines is  sensitive
 to both $b_1$ and $b_2$.  
 It should, in principle,  be possible to
 use observations of the shape dependence of   $\C_3$ to determine
 the bias parameters.  

\begin{figure}
  \includegraphics[width=113mm]{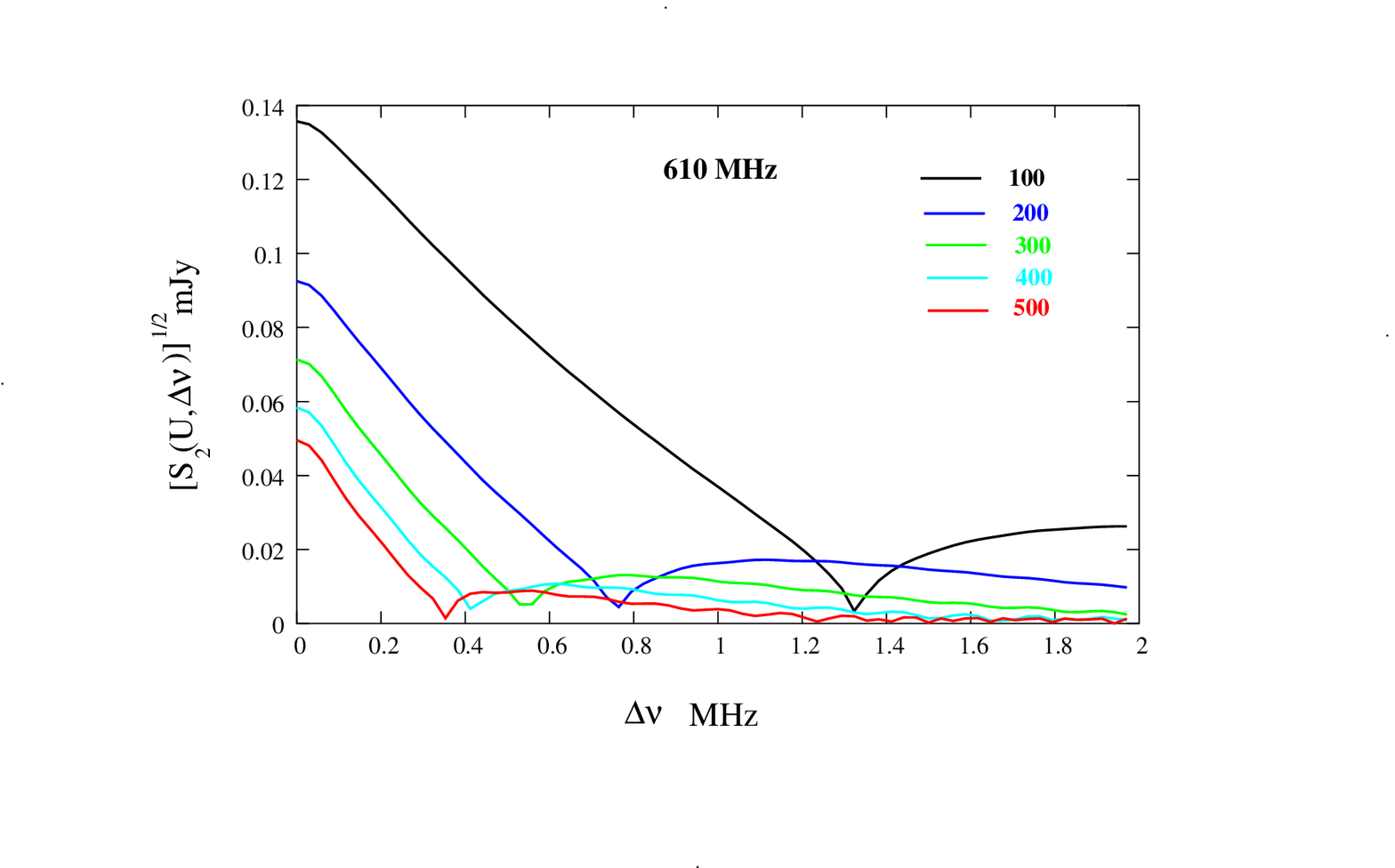}
  \caption{This shows the expected  two visibility correlation signal
  $[\C_2(U,\Delta \nu)]^{1/2}$  as  a function of  $\Delta \nu $ for
  the   different values of $U$ shown in the figure at a  central
  frequency of  $610 {\rm   \,MHz }$}. 
  \label{fig:4}
\end{figure}

\begin{figure}
  \includegraphics[width=84mm]{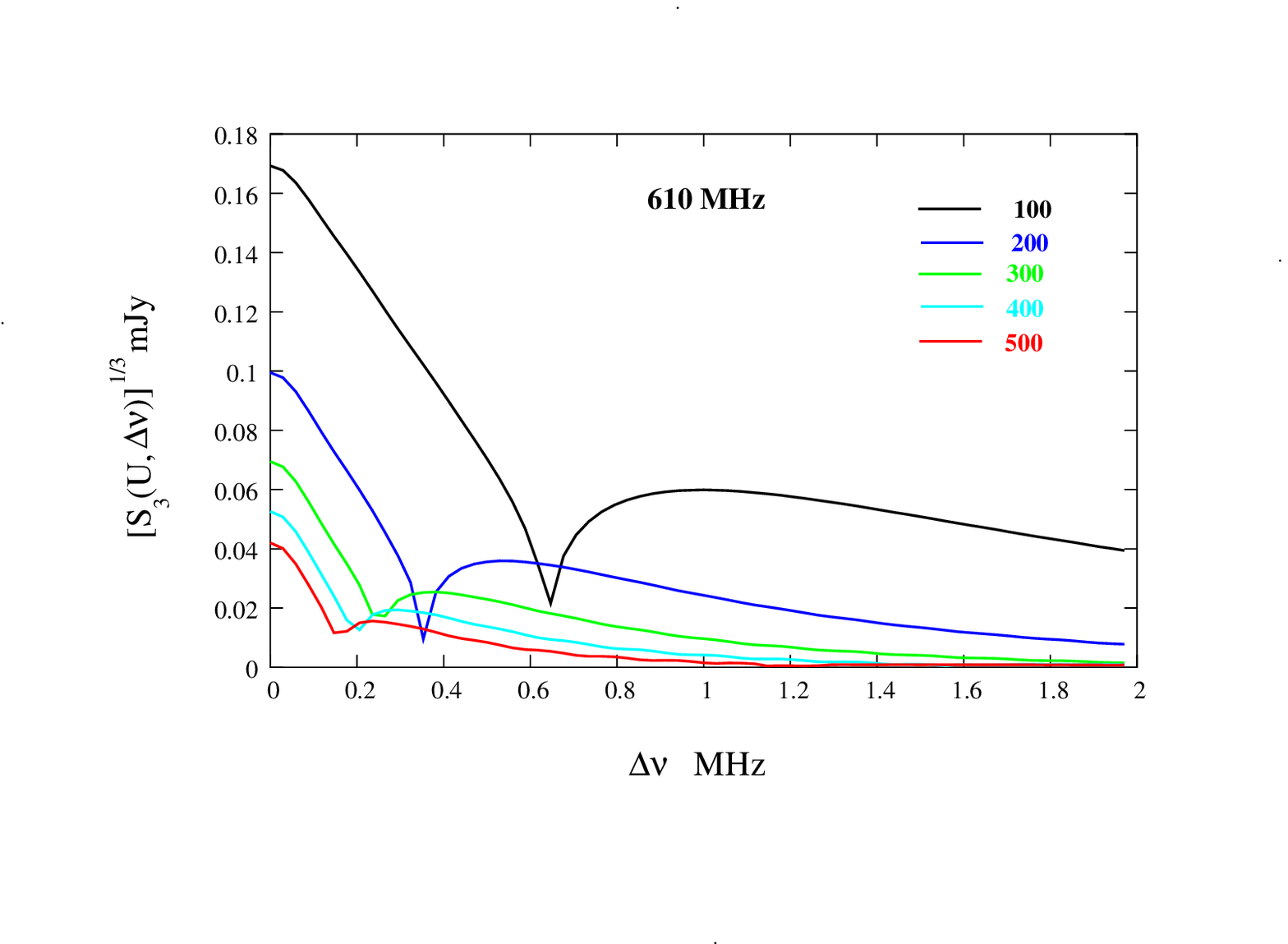}
  \caption{This shows the expected  three visibility correlation signal
  $[\C_3(U,\Delta \nu)]^{1/3}$  as  a function of  $\Delta \nu $ for
  the   different values of $U$ shown in the figure at a  central
  frequency of  $610 {\rm   \,MHz }$}.  
  \label{fig:5}
\end{figure}

We next consider the correlation between visibilities at different
frequencies. The results are presented at fixed values of the
baselines, and Figures \ref{fig:4} and \ref{fig:5} show \,$[\C_2(U,\Delta
  \nu)]^{1/2}$ and $[\C_3(U,\Delta
  \nu)]^{1/3}$ respectively, both at the central frequency of $610
\,{\rm MHz}$. We find that the correlations decay rapidly as the
frequency separation $\Delta \nu$ is increased. We have $\C(U,\delta
\nu) =0$ at $\Delta \nu \sim 1.3 \,{\rm MHz}$ for $U=100$,  and there is
a weak anticorrelation $\C_2 < 0$ for larger $\Delta \nu$. The
visibilities at larger baselines get decorrelated faster with
increasing $U$, and the zero crossing is at $\sim 0.7 \,{\rm
  MHz}$ at $U=100$. The three visibility correlation shows a similar
behavior, the main difference being that $[\C_3(U,\Delta \nu)]^{1/3}$
decays  much faster than $[\C_2(U,\Delta \nu)]^{1/2}$ with increasing 
$\Delta \nu$. It may be noted that the overall behavior is similar at
other values of the central frequency,  with the difference that
$\C_2$ and $\C_3$ decay faster with increasing $\Delta \nu$ at lower
values of the central frequency.

An  important issue which we have not addressed here in any detail is
 the extraction  of the HI signal from various  contaminants which are
 expected to dominate at low frequencies. 
The issue of extracting the redshifted
 21 cm signal has received considerable  attention
 (eg. \citealt{shaver}; 
 \citealt{dimat1} ;  \citealt{oh3};\citealt{morl}  ; \citealt{zald}  ;
 \citealt{cooray} ;  \citealt{sant5} ; \citealt{wang}; \citealt{mor2}).  
The contribution from the continuum sources of contamination at two
 different frequencies is expected to be correlated even at large
 $\Delta \nu$, whereas the HI 
 signal  is found to be uncorrelated beyond something like $1.3 \,
 {\rm MHz}$ or even less depending on the value of $U$. It is then, in
 principle, straightforward to fit the visibility correlations $\C_2$
 and $\C_3$  at large $\Delta \nu$ and remove any slowly  varying
 component thereby separating the contaminants from the HI signal. We
 also use this opportunity to note that this is a major advantage of
 using visibility correlations as compared to the  angular power 
 spectrum  which has been found to have  substantial correlations even
 at two  frequencies separated by $\sim 10 \, {\rm MHz}$
 \citep{sant5}. 

System noise  is  inherent in all   radio interferometric observations. 
 System noise and the  observation time needed to detect the HI signal
 is another important issue.
Here we briefly discuss the signal to noise ratio and approximate integration time required to detect the HI signal for $\C_2$ and $\C_3$. 
 We consider
an array of $N$ antennas, the observations lasting a time duration
$t$, with frequency channels of 
width $\delta \nu$ spanning a total bandwidth $B$.  It should be noted
that the effect of a finite channel width $\delta \nu$ has not been
included in our calculation which assumes infinite frequency
resolution. This effect can be easily included by convolving our
results for the visibility correlation with the frequency
response function of a single channel. Preferably, $\delta \nu$
should be much smaller than the frequency separation at which the
visibility correlation become  uncorrelated.  We use $\cal S $ denote the  
frequency separation within which the visibilities are correlated,  and
beyond which they become uncorrelated.We use $\N_2$
and $\N_3$ to denote the rms. noise in $\C_2$ and $\C_3$. we have assumed that thermal noise are Gaussian random fields. It is well known that  $\N_2 = \left( \frac{2 k_B T_{SYS}}{A_{ef}}   \right)^2 \frac{1}{\delta \nu \, t}$
\citep{thomp}, and we have $\N_3\sim \left( \frac{2 k_B 
  T_{SYS}}{A_{ef}}   \right)^{3} \frac{1}{(\delta \nu \, t)^{3/2}}$, where $T_{SYS}$ is the system temperature and $A_{ef}$ is the effective
area of a single antenna.The noise contribution  will be reduced by a
factor $1/\sqrt{N_o}$ if we  combine $N_o$ independent samples of the 
visibility correlation.  A possible observational strategy for  a
preliminary detection of the HI signal would be to combine the
visibility correlations at all baselines and frequency separations
where there is a reasonable amount of signal. This gives
$N_o=[N(N-1)/2] \, (B/\delta \nu) \, (\cal S/\delta \nu)$ for the two
visibility correlation and $N_o=[N(N-1)(N-2)/6] \,(B/\delta \nu) \,
{(\cal S/\delta \nu)}^2 $ for the three visibility correlations. Combining all of this
we have $[\N_2]^{1/2} \sim  \left( \frac{2 k_B
  T_{SYS}}{A_{ef}}   \right) \left[ \frac{2}{N (N-1) B \cal S }
  \right]^{1/4} \frac{1}{t^{1/2}}$ and $[\N_3]^{1/3} \sim  \left( 
\frac{2 k_B   T_{SYS}}{A_{ef}}   \right) \left[ \frac{6}{N (N-1)(N-2) B\, {\cal S}^2 } \right]^{1/6} \frac{1}{t^{1/2}}$. Using values for the GMRT at $ \sim 610\, {\rm MHz}$, $(2 k_B
T_{SYS}/A_{ef})=144 \, {\rm Jy}$,  $B=16 \, {\rm MHz}$, there are $14$ 
antennas within $U \le 1000$  where the signal is 
strong and $\cal {S}$ = $0.5 \,  {\rm MHz}$ beyond which the signal is
uncorrelated we find that it is possible to achieve noise levels of \, 
$[\N_2]^{1/2} \sim  0.03 \, {\rm mJy}$ and $[\N_3]^{1/3} \sim  0.045 \, {\rm mJy}$
which are below the signal  with  $200 \, {\rm hrs}$ of integration. Our present estimations indicate that $200$ to $1000$ hours of  observation are needed to
 detect the HI signal.

\section{Acknowledgments}
  SSA  would like to thank the
 CSIR, Govt. of India for financial support through  a senior research fellowship. SB would also like to acknowledge BRNS, DAE,
 Govt. of India,for  financial support through sanction
 No. 2002/37/25/BRNS. SKP would like to acknowledge the Associateship
 Program, IUCAA for   supporting his visit to IIT, Kgp and CTS,IIT Kgp
 for the use of its facilities.

\end{document}